\documentclass[aps,amsmath,amssymb,twocolumn,showpacs]{revtex4-1}

\usepackage{graphicx}
\usepackage{dcolumn}
\usepackage{bm}
\usepackage{color}

\begin{document}

\title{Magnetic-field controlled anomalous refraction in doped semiconductors}

\author{E. Moncada-Villa$^{1}$}
\author{A.~I. Fern\'andez-Dom\'{\i}nguez$^{2}$}
\author{J.~C. Cuevas$^{2,3}$}

\affiliation{$^{1}$Escuela de F\'{\i}sica, Universidad Pedag\'ogica y Tecnol\'ogica de Colombia,
Avenida Central del Norte 39-115, Tunja, Colombia}

\affiliation{$^2$Departamento de F\'{\i}sica Te\'orica de la Materia Condensada
and Condensed Matter Physics Center (IFIMAC), Universidad Aut\'onoma de Madrid,
E-28049 Madrid, Spain}

\affiliation{$^{3}$Department of Physics, University of Konstanz, D-78457 Konstanz, Germany}

\date{\today}

\begin{abstract}
We predict here that a slab made of a doped semiconductor can exhibit anomalous refraction under the application
of a static magnetic field. This anomalous refraction takes place in the far-infrared range and it occurs for any angle of
incidence. We show that this effect is due to the fact that a doped semiconductor under a magnetic field can behave,
to some extent, as a hyperbolic metamaterial. We also show that the occurrence of this anomalous refraction enables 
a semiconductor slab under a magnetic field to partially focus the electromagnetic radiation. The remarkable thing 
in our case is that we deal with naturally occurring materials and the anomalous refraction can be tuned at will 
with an external field.
\end{abstract}

\maketitle

\section{Introduction}

Veselago's proposal \cite{Veselago1968} of a negative refraction index triggered the design and fabrication
of artificial materials, known as metamaterials, with peculiar optical properties \cite{Shalaev2007}.
In principle, such materials were meant to possess an isotropic optical response, that is, their dielectric permittivity
and magnetic permeability are given by scalar functions, being both negative in a certain frequency range. However,
the experimental realization of these materials is usually carried out with an ensemble of building blocks, leading to anisotropic
effective optical responses \cite{Pendry2004,Han2008}. A notable example of materials with engineered anisotropy 
are the hyperbolic metamaterials, which are uniaxial media described by a permittivity that is a diagonal tensor of the
form: $\hat \epsilon = \mbox{diag}(\epsilon_{xx},\epsilon_{xx},\epsilon_{zz})$ \cite{Poddubny2013,Ferrari2015}. These metamaterials
are usually fabricated with the help of metal-dielectric multilayered structures or nanowire arrays \cite{Poddubny2013,Ferrari2015}.
What makes special this class of metamaterials is the fact that for certain frequency ranges it holds that the real part of
$\epsilon_{xx}$ and $\epsilon_{zz}$ have opposite signs. This, in turn, leads to the fact that the dispersion relation of
the electromagnetic waves propagating inside these materials is hyperbolic, i.e., it adopts the following form
\begin{equation}
\frac{k_x^2}{\epsilon_{zz}} + \frac{k_z^2}{\epsilon_{xx}} = \frac{\omega^2}{c^2},
\end{equation}
where $\omega$ is the frequency, $c$ is the speed of light, and $k_{x,z}$ are the corresponding components of the wave vector.
As a consequence, these metamaterials are able to convert incoming evanescent waves into propagating ones. This special feature
of hyperbolic metamaterials makes it possible their application for negative refraction devices, subwavelenght imaging and
superlensing, and hyperbolic waveguides \cite{Poddubny2013,Ferrari2015}. In addition, they can exhibit
anomalous refraction, which is the inversion of the transversal component of the Poynting vector once the electromagnetic wave
enters into the medium \cite{Hu2002}.

In this work we explore the possibility to use a simple semiconductor to mimic the behavior of a hyperbolic metamaterial
in the far-infrared range of the electromagnetic spectrum. In particular, we study the possibility to induce an anomalous 
refraction by simply applying a static magnetic field to a doped semiconductor. This idea was inspired by our recent work 
where we showed that, under certain circumstances, doped semiconductors can behave as hyperbolic near-field thermal emitters 
when subjected to an external magnetic field \cite{Moncada2015}. This led us to the question of whether semiconductors under 
magnetic fields can exhibit the defining properties of hyperbolic metamaterials in other contexts. As we show below,
this is actually the case and we shall illustrate here this phenomenon with the case of a slab made of $n$-doped indium antimonide (InSb),
a semiconductor that has been amply studied in the context of magneto-plasmonics \cite{Palik1976}, near-field thermal radiation
\cite{Moncada2015,Ben-Abdallah2016,Zhu2016,Abraham-Ekeroth2018}, magnetically controlled subwavelength resolution \cite{Cheng2015},
and in biosensing plataforms \cite{Sreekanth2016}. To be more precise, we show in this work that a slab of InSb in the presence
of a magnetic field can exhibit anomalous refraction for any angle of incidence in a wide range of frequencies in the far-infrared
We also predict that the occurrence of this anomalous refraction partially focus the 
electromagnetic radiation. The notable thing about this discovery is that this anomalous refraction only involves the use of 
a naturally occurring material and it is highly tunable with an external field.

The rest of the paper is organized as follows. In Sec.~\ref{anom_refr}, we present the system under study and briefly
remind the basics of anomalous refraction. Section~\ref{anom_refr_insb} is devoted to the description of the main results
of the work. In particular, we present a detailed analysis of the anomalous refraction in a InSb slab subjected to a
static magnetic field. In Sec.~\ref{sec-focusing} we explore the possibility of inducing partial focusing in a
InSb slab with the application of an external magnetic field. The main conclusions of this work are briefly summarized 
in Sec.~\ref{conclusions}. Finally, we present additional results in Appendix A to clarify some of 
the statements made in different sections.


\section{System and reminder of anomalous refraction}\label{anom_refr}

\begin{figure}[t]
\includegraphics[width=0.6\columnwidth,clip]{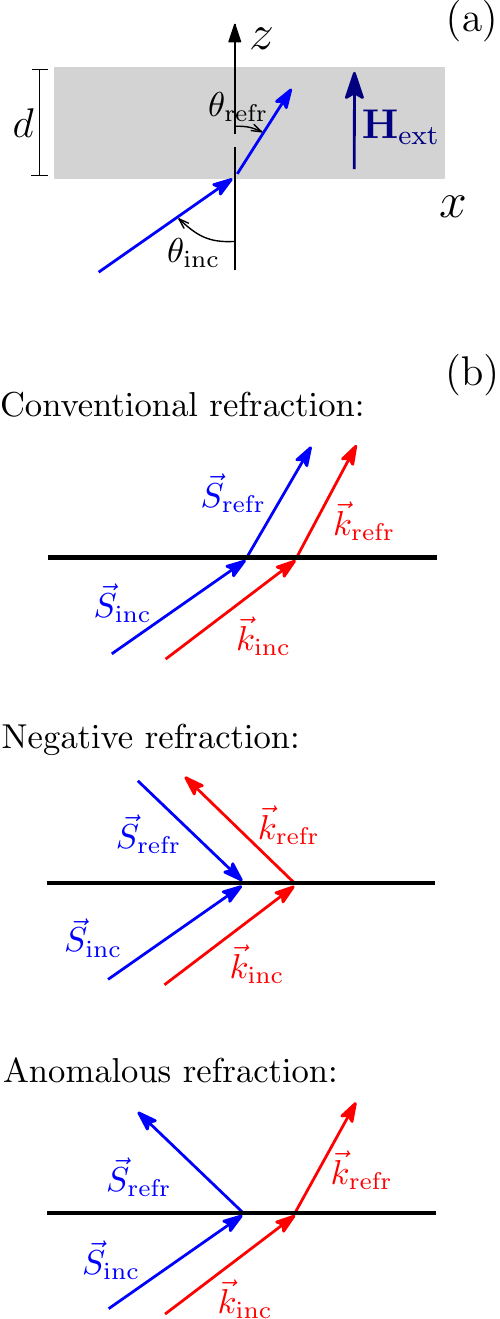}
\caption{(color online) (a) Plane wave impinging (with an angle of incidence $\theta_{\rm{inc}}$) in a planar slab,
which is under the action of an external magnetic field that is perpendicular to the slab. The refraction angle
$\theta_{\rm{refr}}$ stands for the direction of the Poynting vector once the wave enters into the slab.
(b) Schematic representation of different types of refraction that can take place at the interface: conventional,
negative, and anomalous.}
\label{fig_slab}
\end{figure}

The system that we consider is a slab of thickness $d$ made of a doped semiconductor, see Fig.~\ref{fig_slab}(a).
The slab is surrounded by air (or vacuum) and we assume that there is a static magnetic field applied perpendicular
to the slab's surfaces, which is the $z$-direction according to Fig.~\ref{fig_slab}(a), i.e.,  $\mathbf H_{\rm ext} =
H\mathbf{\hat{z}}$. Because of the existence of free carriers in a doped semiconductor, it is well-known that
the application of a magnetic field induces a magneto-optical activity that in our geometry can be generically
described by the following permittivity tensor \cite{Palik1976,Zvezdin1997}
\begin{equation}
\label{perm-tensor-xy}
\hat \epsilon = \left( \begin{array}{ccc}
 \epsilon_{xx} & \epsilon_{xy} &0 \\
-\epsilon_{xy} & \epsilon_{xx} & 0 \\
0              & 0             & \epsilon_{zz} \end{array} \right).
\end{equation}
The different elements in this tensor depend on the applied magnetic field and their exact form will be discussed
in the next section. For the moment, the important thing to notice is that the diagonal part of this tensor has
exactly the same form as in the case of a hyperbolic material. However, this tensor also has off-diagonal elements,
which are absent in a uniaxial material and are induced by the applied field. These off-diagonal elements are responsible
for the polarization conversion that an electromagnetic wave undergoes when it is reflected by the slab or transmitted
through it. This is actually the basis of the Kerr and Faraday magneto-optical effects that are observed in these materials
\cite{Zvezdin1997}.

Our goal in this work is to analyze the transmission of a plane wave that impinges in this magneto-optical slab with
a certain angle of incidence, $\theta_{\rm inc}$. Here, we assume that the $xz$-plane coincides with the incidence plane.
In particular, we want to find out whether the anomalous refraction can take place in the system under consideration,
something that is not trivial a priori because of the presence of the off-diagonal terms in the permittivity tensor of
Eq.~(\ref{perm-tensor-xy}). This problem will be analyzed in detail in the following section, but for didactic reasons
we are going to assume in this section that we can ignore the off-diagonal elements in Eq.~(\ref{perm-tensor-xy}). This
will allow us to remind the basics of the anomalous refraction and to explain how it comes about.

With the assumption that the permittivity tensor is diagonal, $\hat \epsilon = \mbox{diag}(\epsilon_{xx},\epsilon_{xx},\epsilon_{zz})$,
it is easy to analyze the refraction of a plane wave by the slab \cite{Hu2002}. Considering first the case of transversal
electric (TE) polarized waves $\mathbf{E} = E_0 \, \mathbf{\hat{y}} \, e^{i(k_x x+k_z z - \omega t)}$, the solution of
Maxwell's equations lead to the following dispersion relation inside the slab
\begin{equation}
\label{disp_te_2}
\frac{k_x^2}{\epsilon_{xx}} + \frac{k_z^2}{\epsilon_{xx}} = \frac{\omega^2}{c^2}
\end{equation}
while the corresponding $x$-component of the Poynting vector evaluated right inside the slab (after crossing the first
interface) is given by
\begin{equation}
\label{sx_te_2}
S_x =\frac{k_x |E_0|^2}{2\omega\mu_0},
\end{equation}
where $\mu_0$ is the vacuum permeability. Similarly, for transversal magnetic (TM) waves $\mathbf{H} = H_0 \,
\mathbf{\hat{y}}\, e^{i(k_x x+k_z z-\omega t)}$, one obtains the dispersion relation
\begin{equation}
\label{disp_tm_2}
\frac{k_x^2}{\epsilon_{zz}} + \frac{k_z^2}{\epsilon_{xx}} = \frac{\omega^2}{c^2}
\end{equation}
and the $x$-component of the Poynting vector right inside the slab
\begin{equation}
\label{sx_tm_2}
S_x =\frac{k_x|H_0|^2}{2\omega\epsilon_0}{ \rm{Re}}\left\lbrace \frac{1}{\epsilon_{zz}} \right\rbrace,
\end{equation}
where $\epsilon_0$ is the vacuum permittivity. The condition for the anomalous refraction is a negative $x$-component
of the Poynting vector in the slab [c.f.\ Fig.~\ref{fig_slab}(b)]. From Eq.~\eqref{sx_te_2}, it is clear that is not
possible to have anomalous refraction for an incident TE-polarized wave, while for TM-polarized waves this is possible
provided that $\mbox{Re} \{\epsilon_{zz} \} <0$ [see Eq.~\eqref{sx_tm_2}]. If this condition is fulfilled, the dispersion
relation of Eq.~\eqref{disp_tm_2} requires $\mbox{Re} \{ \epsilon_{xx} \} >0$ in order to have propagating
waves, instead of evanescent ones, inside the slab, as it can be seen from the expression for the $z$-component
of the wave vector:
\begin{equation}
\label{kz}
k_z = \sqrt{\epsilon_{xx}\frac{\omega^2}{c^2}\left(1-\frac{\sin^2\theta_{\rm inc}}{\epsilon_{zz}}\right)}.
\end{equation}
Notice that according to this expression, the anomalous refraction within the uniaxial approximation
occurs for any angle of incidence.
This absence of a critical angle in this slab made of natural occurring MO materials differs from the behavior of
uniaxial bianisotropic (artificial) metamaterials with permittivity $\hat\epsilon = \mbox{diag}(\epsilon_{xx},
\epsilon_{xx},\epsilon_{zz})$ and permeability $\hat\mu = \mbox{diag}(\mu_{xx},\mu_{xx},\mu_{zz})$ \cite{Hu2002}.

Let us stress that the whole discussion above was based on the uniaxial approximation (with no 
off-diagonal elements of the permittivity tensor). In the next section, we investigate how these conclusions are 
modified by analyzing the exact results taking into account the full structure of the permittivity tensor in 
Eq.~(\ref{perm-tensor-xy}). We can anticipate that the conditions for the appearance of the anomalous refraction 
derived above remain valid in the case of a doped semiconductor under a magnetic field. This is actually due to 
the fact that the dispersion relations of the different propagating modes, see Eqs.~(\ref{disp_te_2}) and 
(\ref{disp_tm_2}), are not strongly modified by the off-diagonal elements. This is discussed in some detail in 
Appendix A. Let us also say that, on the contrary, the wave propagation inside the slab is clearly altered by the 
off-diagonal elements.

\section{Anomalous refraction in a doped semiconductor}\label{anom_refr_insb}

\begin{figure}[t]
\includegraphics[width=0.9\columnwidth,clip]{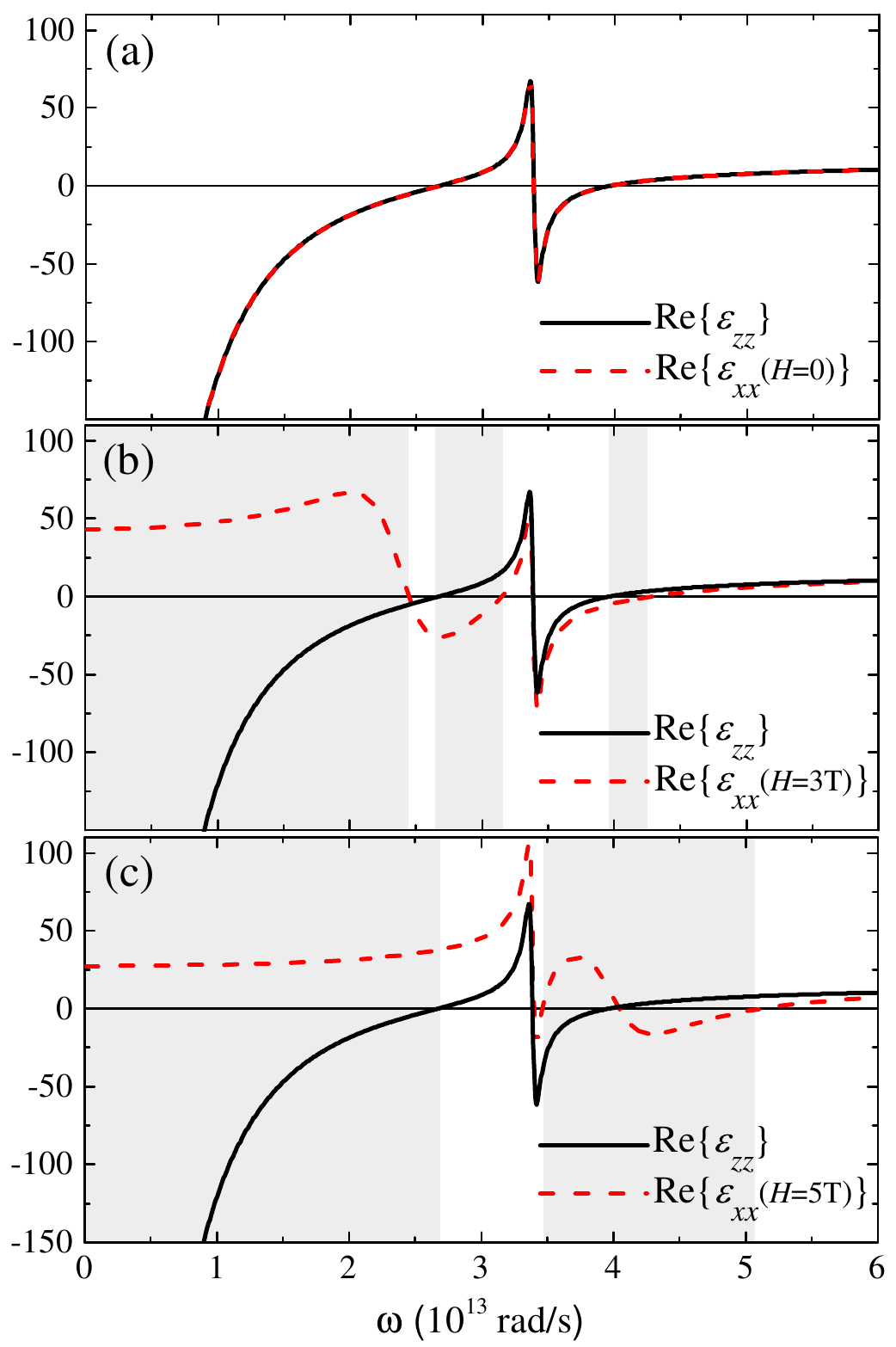}
\caption{(color online) Real part of diagonal elements of the permittivity tensor of Eq.~\eqref{perm-tensor-xy}
as a function of the frequency for several values of external magnetic field $H$: (a) 0 T, (b) 3 T, and
(c) 5 T. Shaded regions correspond to cases where hyperbolic modes exist inside the slab, i.e., regions
where the real parts of $\epsilon_{xx}$ and $\epsilon_{zz}$ have opposite signs.}
\label{fig_epsilon}
\end{figure}
\begin{figure*}[t]
\includegraphics[width=0.8\textwidth,clip]{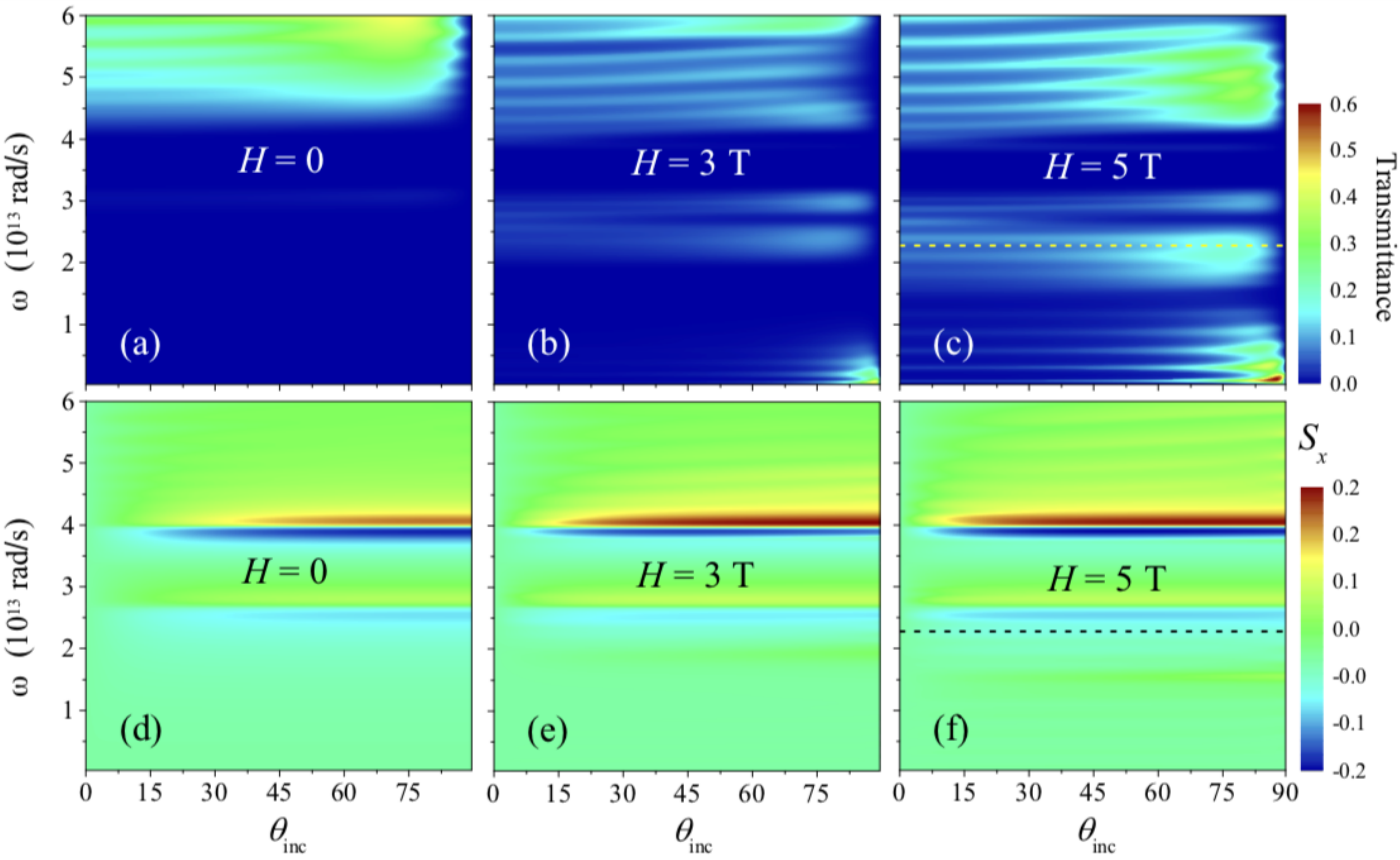}
\caption{(color online) Transmittance as a function of the frequency and the angle of incidence of a
TM-polarized plane wave impinging in a 50 $\mu$m InSb slab that is subjected to an external magnetic field
of (a) 0 T, (b) 3 T, and (c) 5 T. The $x$-component of the corresponding Poynting vector is shown in
panels (d), (e) and (f). The Poynting vector is evaluated inside the slab, right at the interface in which
the plane wave impinges.}
\label{fig_trans_H}
\end{figure*}

In the previous section we have reminded that a hyperbolic metamaterial with a scalar permeability exhibits an
anomalous refraction for any angle of incidence. Since a doped semiconductor is not quite a uniaxial material, see
Eq.~(\ref{perm-tensor-xy}), it is not obvious a priori whether it can also exhibit this peculiar phenomenon.
In order to clarify this issue, we investigate in this section the refraction through a slab made of $n$-doped InSb.
The optical properties of this material under a perpendicular magnetic field, see Fig.~\ref{fig_slab}(a), are
described by a permittivity tensor of the form of Eq.~(\ref{perm-tensor-xy}), where the different elements
of that tensor are given by \cite{Palik1976}
\begin{eqnarray}
\epsilon_{xx}(H) & = & \epsilon_{\infty} \left( 1 + \frac{\omega^2_L - \omega^2_T}{\omega^2_T -
\omega^2 - i \Gamma \omega} + \frac{\omega^2_p (\omega + i \gamma)}{\omega [\omega^2_c -
(\omega + i \gamma)^2]} \right) , \nonumber \\
\label{eq-epsilons}
\epsilon_{xy}(H) & = & \frac{-i\epsilon_{\infty} \omega^2_p \omega_c}{\omega [(\omega + i \gamma)^2 -
\omega^2_c]} , \label{eq-tensor-el} \\
\epsilon_{zz} & = & \epsilon_{\infty} \left( 1 + \frac{\omega^2_L - \omega^2_T}{\omega^2_T -
\omega^2 - i \Gamma \omega} - \frac{\omega^2_p}{\omega (\omega + i \gamma)} \right) . \nonumber
\end{eqnarray}
Here, $\epsilon_{\infty}$ is the high-frequency dielectric constant, $\omega_L$ is the longitudinal
optical phonon frequency, $\omega_T$ is the transverse optical phonon frequency, $\omega^2_p =
ne^2/(m^{\ast} \epsilon_0 \epsilon_{\infty})$ defines the plasma frequency of free carriers of density
$n$ and effective mass $m^{\ast}$, $\Gamma$ is the phonon damping constant, and $\gamma$ is the
free-carrier damping constant. Finally, the magnetic field enters in these expressions via the cyclotron
frequency $\omega_c = eH/m^{\ast}$. There are several features in the previous expressions that are worth
stressing. First, the magnetic field induces an optical anisotropy via the modification of the diagonal
elements. Second, there are two major contributions to the diagonal components of the permittivity tensor:
optical phonons and free carriers. Third, the magnetic field also induces a magneto-optical activity via
the introduction of off-diagonal elements. Fourth, the optical anisotropy is introduced via the free
carriers, which illustrates the need to deal with doped semiconductors. In what follows, and for the
sake of concreteness, we shall concentrate in a particular case taken from Ref.~[\onlinecite{Palik1976}],
where $\epsilon_{\infty} = 15.7$, $\omega_L = 3.62 \times 10^{13}$ rad/s, $\omega_T = 3.39\times 10^{13}$
rad/s, $\Gamma = 5.65 \times 10^{11}$ rad/s, $\gamma = 3.39 \times 10^{12}$ rad/s, $n = 1.07 \times 10^{17}$
cm$^{-3}$, $m^{\ast}/m = 0.022$, and $\omega_p = 3.14 \times 10^{13}$ rad/s. As a reference, let us
say that with these parameters $\omega_c = 8.02 \times 10^{12}$ rad/s for $H=1$ T.

In Fig.~\ref{fig_epsilon} we show the real part of the permittivity tensor elements $\epsilon_{xx}$ and
$\epsilon_{zz}$ for different values of the magnitude of the external magnetic field. As one can see in panels (b)
and (c), as the external magnetic field increases, there appear broader frequency regions in which the
real parts of $\epsilon_{xx}$ and $\epsilon_{zz}$ have opposite signs, which is the necessary condition
for the hyperbolic modes to exist. In those regions, and as explained in the previous section, the uniaxial
approximation predicts that incoming waves may be refracted into the slab as hyperbolic waves with $k_z$
given by the expression of Eq.~\eqref{kz}.

Now we turn to the analysis of the refraction of a plane wave impinging in our InSb slab. In particular,
we focus on the case of TM-polarized plane waves since, as we discussed in the previous section, these
are the ones that can undergo anomalous refraction. The required calculations for the analysis of this
phenomenon are relatively straightforward and can be carried out with standard methods of multilayer
systems made of optically anisotropic materials \cite{Yeh1988}. In our case, we have made use of the 
scattering-matrix method of Ref.~\cite{Caballero2012}. Let us emphasize that from now on, 
unless stated otherwise, we present numerically exact results obtained using the full structure of 
the permittivity tensor, including the off-diagonal elements and the real and imaginary parts of the 
different elements, as described in Eq.~(\ref{eq-tensor-el}). In Fig.~\ref{fig_trans_H}(a-c)
we show the transmittance through a 50 $\mu$m-thick InSb slab as a function of the frequency and the angle
of incidence for a TM-polarized plane wave and several values of the magnetic field. Notice that the field
changes significantly the transmittance through the slab in several frequency regions. 

\begin{figure*}[t]
\includegraphics[width=0.9\textwidth,clip]{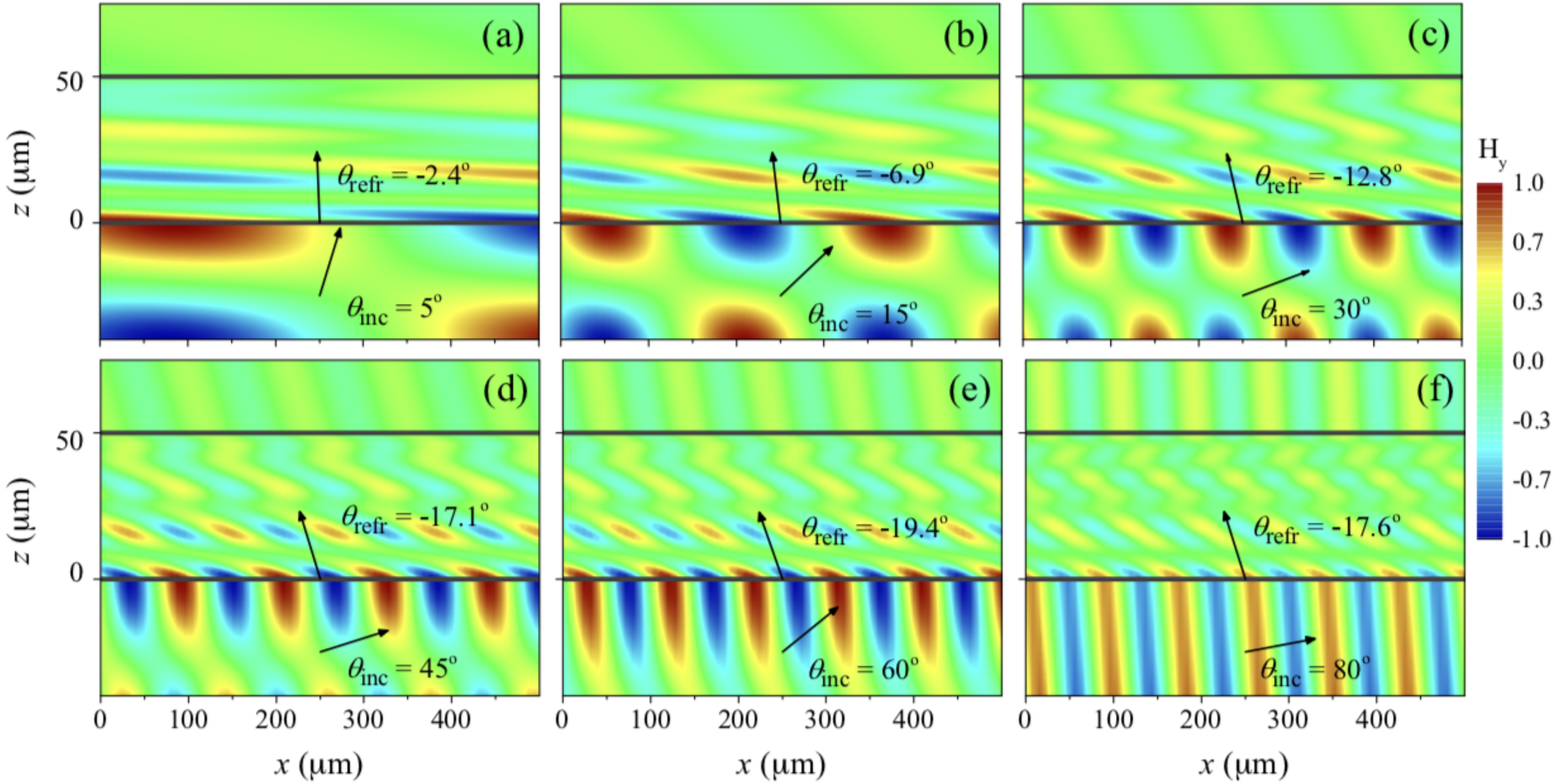}
\caption{(color online) Spatial distribution of the $y$-component of the magnetic field of an incoming TM-polarized
plane wave with frequency $2.28\times 10^{13}$ rad/s that impinges in a 50 $\mu$m-thick InSb slab under the action
of an external magnetic field of 5 T [dashed line in Fig.~\ref{fig_trans_H}(c,f)]. The different panels correspond
to different values of the incidence angle $\theta_{\rm{inc}}$. The angle $\theta_{\rm{refr}}$ indicates the direction
of the Poynting vector immediately after the wave crosses the first interface. }
\label{fig_field_insb}
\end{figure*}

In order to determine whether the anomalous refraction may take place in this system, we also show in
Fig.~\ref{fig_trans_H}(d-f) the $x$-component of the Poynting vector for the same cases as in panels (a-c). Let 
us stress that in this figure we evaluate the Poynting vector inside the slab right at the interface in which 
the plane wave impinges. As one can see, the $x$-component of the Poynting vector is actually negative for some 
frequencies and when this occurs, it remains negative for any angle of incidence. It is important to stress that 
these frequency regions are in all cases regions where the real parts of $\epsilon_{xx}$ and $\epsilon_{zz}$ have
opposite signs, i.e., they correspond to regions where the hyperbolic modes exist (at least in the
uniaxial approximation). These results demonstrates that indeed a slab of a doped semiconductor can exhibit
a magnetic-field induced and controlled anomalous refraction.

In order to get a deeper insight into how this anomalous refraction takes place, we now analyze in more detail
its angle dependence. For this purpose, we present in Fig.~\ref{fig_field_insb} results for spatial distribution
of the $y$-component of the magnetic field when a TM-polarized plane wave impinges on a 50 $\mu$m-thick slab at
different angles of incidence. In this case, we have fixed the magnitude of the external field to 5 T and we have
chosen a frequency of $2.28\times 10^{13}$ rad/s. With these parameters, the hyperbolic condition is fulfilled,
i.e., the real parts of $\epsilon_{xx}$ and $\epsilon_{zz}$ have opposite signs, see Fig.~\ref{fig_epsilon} and
also Fig.~\ref{fig_trans_H}(c,f). In the different panels of Fig.~\ref{fig_field_insb} we provide the result for
the angle $\theta_{\rm{refr}}$ that indicates the direction of the Poynting vector immediately after the wave crosses
the first interface. Notice that irrespective of the angle of incidence, the angle $\theta_{\rm{refr}}$ is always
negative, which indicates that an anomalous refraction takes place. This fact confirms, in turn, the predictions made
in the previous section based on the uniaxial approximation, namely the fact that an external magnetic field may cause 
the anomalous refraction of an incident wave immediately after it enters into the slab, as well as the absence of a 
critical angle.

\begin{figure}[t]
\includegraphics[width=0.9\columnwidth,clip]{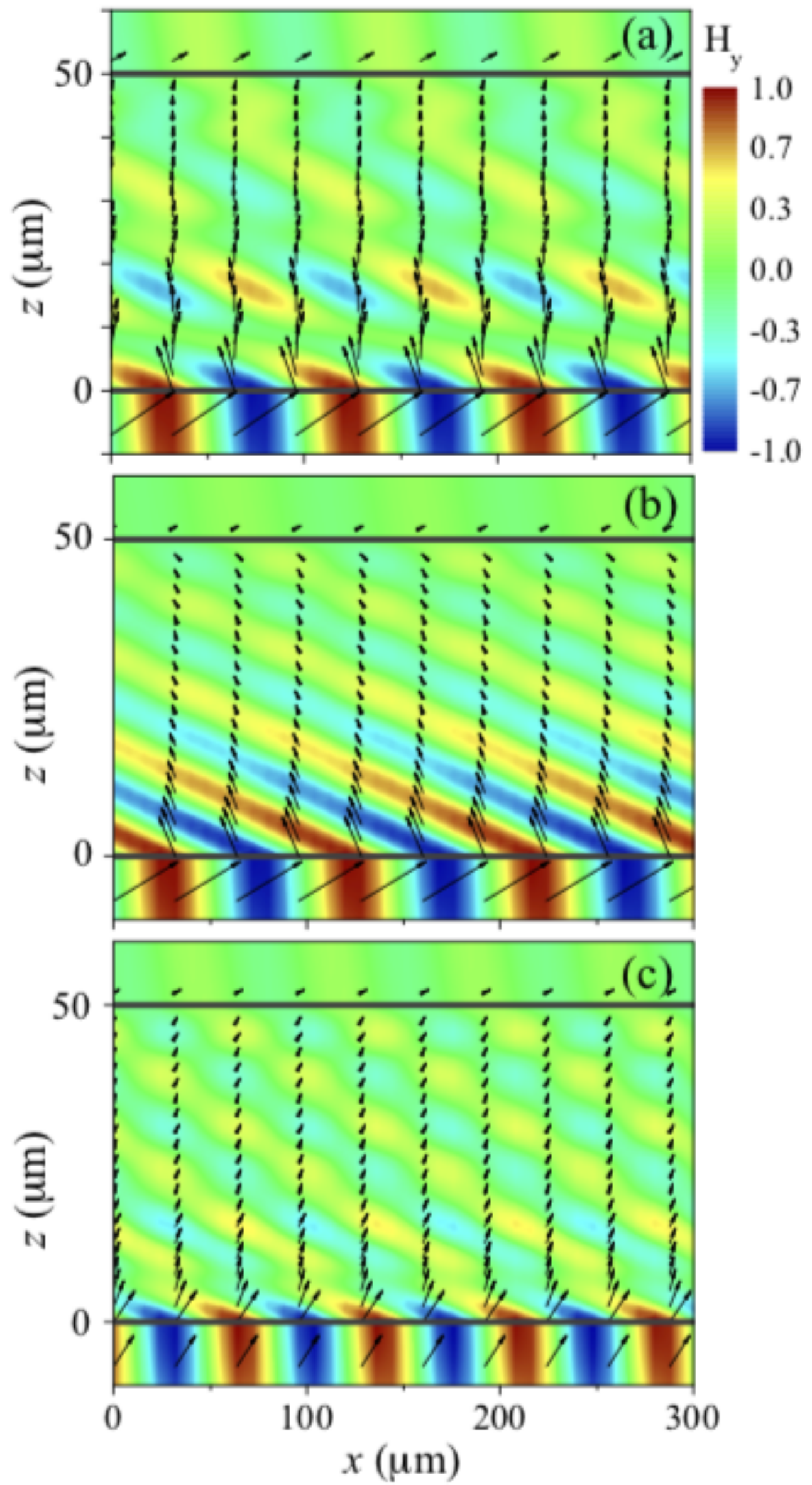}
\caption{(color online) (a) Poynting vector of an incoming wave upon a 50 $\mu$m-thick InSb slab under the action of
an external magnetic field of 5 T with an angle of incidence of $\theta_{\rm{inc}}=60^{\rm o}$ and a frequency of
2.28$\times 10^{13}$ rad/s. In this case $\epsilon_{xx}>0$ and $\epsilon_{zz}<0$. The calculation has been
done taking into account the full structure of the permittivity tensor. (b) The same as in panel (a), but assuming
that there is no magneto-optical activity ($\epsilon_{xy} =0)$. (c) The same as in panel (a), but for a
frequency of $3.0 \times 10^{13}$ rad/s for which the real parts of $\epsilon_{xx}$ and $\epsilon_{zz}$
are both positive.}
\label{fig_poynting}
\end{figure}

As we have pointed out several times already, a doped semiconductor under an external magnetic field is not
exactly equivalent to a hyperbolic material. The external magnetic field also induces off-diagonal elements
of the permittivity tensor that are responsible for the polarization conversion and the corresponding
magneto-optical effects. Moreover, these elements modify the dispersion relation of the propagating 
waves inside the slab, as we briefly discuss in Appendix A. In this sense, one may wonder what their role 
is in this anomalous refraction. To
answer this question, we present in Fig.~\ref{fig_poynting}(a,b) a comparison of the results for the spatial
distribution of the Poynting vector in a case where the anomalous refraction takes place, which were computed
taking into account the full structure of the permittivity tensor ($\epsilon_{xy}\neq0$), panel (a), and
with the uniaxial approximation ($\epsilon_{xy} = 0$), panel (b). The different parameters are indicated
in the caption of that figure. 

For the sake of comparison, we show in Fig.~\ref{fig_poynting}(c)
the exact results for the Poynting vector in the case of a frequency where there are no hyperbolic modes. As one
can see in Fig.~\ref{fig_poynting}(a) for the exact results, the $x$-component of the Poynting vector inside the
slab oscillates between negative (anomalous refraction) and positive (conventional refraction) values, which is
due to the polarization conversion that occurs inside the slab due to the magneto-optical activity. This is
in marked contrast with the results shown in Fig.~\ref{fig_poynting}(b) that were computed with the uniaxial
approximation and therefore, there is no polarization conversion. In this case, the anomalous refraction survives
throughout the slab, with the magnitude of the Poynting vector decaying inside the slab simply due to the
absorption in this material. These results illustrate the basic difference in the anomalous refraction between
semiconductors under a magnetic field and ideal hyperbolic materials. Let us say to conclude this section that
Fig.~\ref{fig_poynting}(c) illustrates the fact that in the more conventional case when the real parts of
$\epsilon_{xx}$ and $\epsilon_{zz}$ have the same sign, then the $x$-component of the Pointing vector
only takes positive values, irrespective of the fact that there is some degree of polarization conversion due
to the fact that $\epsilon_{xy} \neq 0$.

\begin{figure}[t]
\includegraphics[width=\columnwidth,clip]{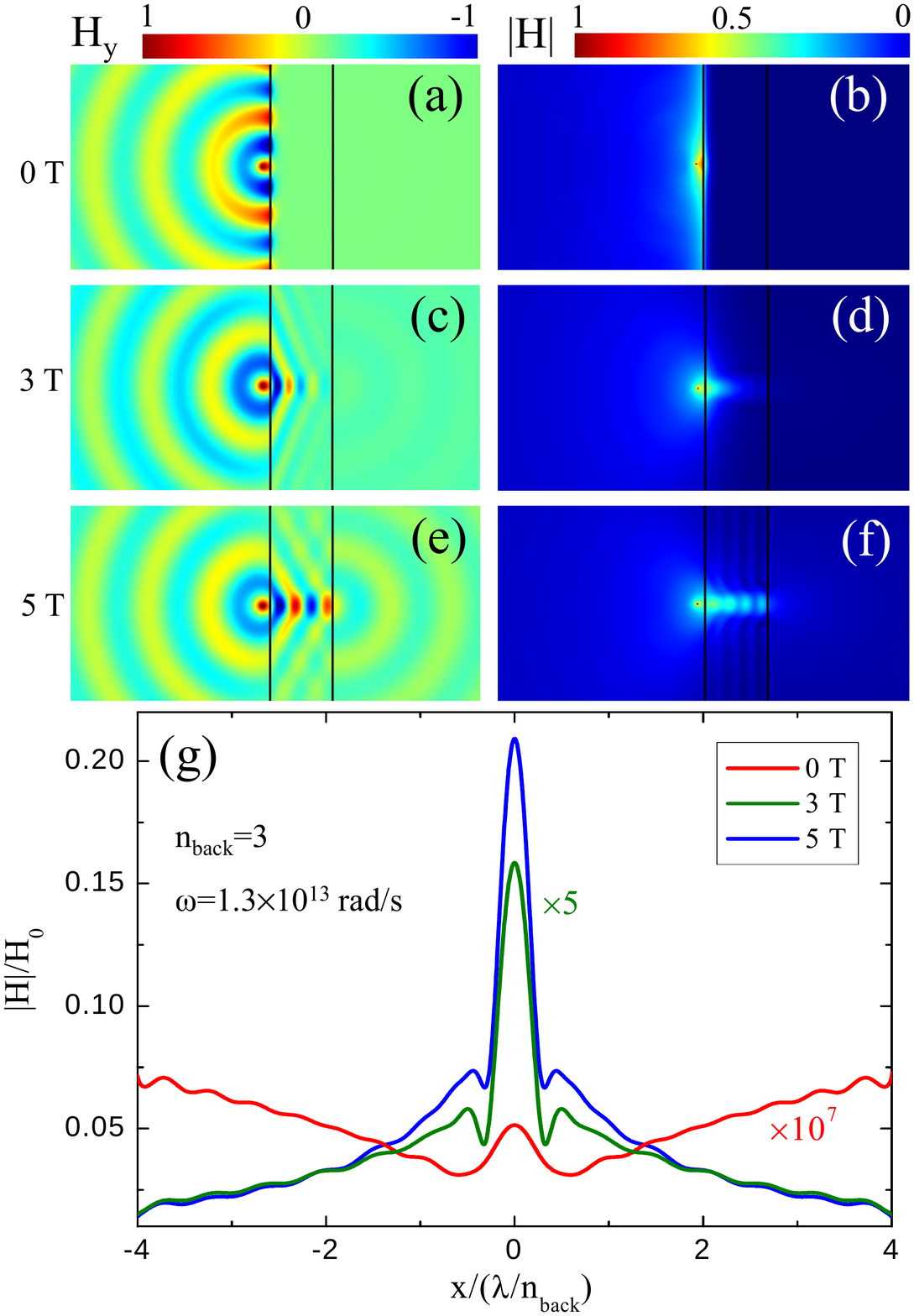}
\caption{(color online) (a) The distribution of the $y$-component of the magnetic field in a simulation of 
the focusing in the absence of an external magnetic field. In this simulation a point radiation source is 
located at a distance of 6~$\mu$m from a InSb slab of thickness 50 $\mu$m. The slab is embedded by a dielectric
medium with a refractive index equal to $n_{\rm back} = 3$ and the radiation frequency is equal to 
$\omega = 1.3 \times 10^{13}$ rad/s. (b) The same thing as in panel (a), but in this case we show the 
magnitude of the magnetic field. (c-d) The same as in panels (a-b), but for an external static field of 3 T. 
(e-f) The same as in panels (a-b), but for an external static field of 5 T. (g) Amplitude of the transmitted 
magnetic field normalized to $H_0$ (source field) evaluated at the surface of the slab.} 
\label{fig_focusing}
\end{figure}

\section{Partial focusing} \label{sec-focusing}

The occurrence of anomalous refraction in doped semiconductors under a static magnetic field naturally raises 
the question of whether other related phenomena can also take place in these systems. For instance, it is 
well-known that the anomalous refraction in hyperbolic metamaterials enables the realization of lenses 
analogue to the superlens made from negative-index metamaterials \cite{Poddubny2013}. A basic difference, 
however, is the fact that while in negative-index materials the group and phase velocities are always 
antiparallel, in the case of hyperbolic metamaterials the relative orientation of those two velocities 
depends on the propagation direction relative to the principal axis. This fact results in partial focusing 
of radiation by a hyperbolic metamaterial slab \cite{Smith2003,Fang2009,Liu2009,Li2011}, a phenomenon that has
been observed for microwaves \cite{Smith2004}. In this respect, the goal of this section is to explore the 
possibility of observing partial focusing in slabs made of doped semiconductors under an external magnetic field.

For this purpose, we have carried out a series of finite-element simulations exploring the occurrence of 
focusing phenomena in the 50~$\mu$m thick doped InSb slabs studied in the previous section under plane wave 
illumination. The calculations were performed using the frequency-domain Maxwell's Equation solver implemented
in COMSOL MULTIPHYSICS$^{\rm TM}$, the frequency of operation was set to $1.3\times10^{13}$ rad/s, and the 
tunability of the response of the system against the external static magnetic field was analyzed. The external 
excitation was a point-like magnetic field source of the form $\mathbf{H_0}=H_0\mathbf{\hat{y}}$, located
$6~\mu$m away from the doped InSb surface. In order to reduce the impedance mismatch, enhance transmission, 
and improve the visibility of the magnetic fields, we set the background refractive index to $n_{\rm back}=3$, 
which is a good approximation for bulk semiconductors such as GaN or ZnS in the far-infrared range~\cite{Han2007}.

Figure~\ref{fig_focusing} shows the $y$-component (left panels) and magnitude (right panels) of the magnetic fields 
for three different amplitudes of the static field $H$: 0 T (a-b), 3 T (c-d) and 5 T (e-f). The amplitude in all 
field maps are normalized to 1. We can observe that in the absence of static fields, the slab is completely opaque, 
as expected from Fig.~\ref{fig_epsilon}(a), and it behaves as a perfect conductor (with very large, negative and 
isotropic permittivity). Applying an external magnetostatic field, the transmission increases, see Fig.~\ref{fig_trans_H}, 
as the permittivity becomes hyperbolic. At 3 and 5 T, the transmitted fields concentrate at the output, in
the near-field of the slab, the signature of the focusing effect we were seeking for. Figure~\ref{fig_focusing}(g) 
displays the amplitude of the transmitted magnetic field normalized to $H_0$ evaluated at the surface of the slab, 
showing that despite the absence of curvature in the slab, it indeed acts as a lens, able to yield high-resolution 
images of the excitation source. We note that the plots in Fig.~\ref{fig_focusing} were obtained through 2D
calculations assuming translational symmetry in the fields along $y$-direction. The off-diagonal terms in the 
permittivity were not accounted for in these simulations. In order to test their impact in the partially-focused 
field profiles, we also performed more coarsely meshed full 3D calculations and they are discussed in Appendix A. 
Those results allowed us to conclude that a non-vanishing $\epsilon_{xy}$ only reduce the transmissivity of the 
slab, without diminishing its lensing performance.

\section{Conclusions}\label{conclusions}

In summary, we have demonstrated in this work that the application of an external magnetic field may induce 
anomalous refraction at the interface between a dielectric media and a doped semiconductor. We have 
illustrated this phenomenon with the example of InSb, a polar semiconductor, but we anticipate that
this phenomenon can take place in a great variety of semiconductors, including non-polar ones such as Si. 
The anomalous refraction and focusing effect reported in this work occurs in the infrared region of the 
electromagnetic spectra and it is not restricted by the angle of incidence. We have also discussed the
similarities with the anomalous refraction that takes place in hyperbolic metamaterials and the differences 
due to the magneto-optical activity of semiconductors under a static magnetic field. Moreover, we have 
shown that the occurrence of this anomalous refraction enables planar semiconductor slabs to partially 
focus the electromagnetic radiation. A salient feature of the phenomenona predicted in this work
is that, contrary to what happens in hyperbolic metamaterials, they only require naturally occurring 
materials and they can be tuned at will with an external field. We think that our work can trigger
off the realization of experiments aiming at the confirmation of our predictions and it can also stimulate 
theoretical work investigating related physical phenomena such as subwavelength imaging or magnetic-field 
controlled wave-guiding.

\section*{Acknowledgments}
We thank A. Garc\'{\i}a-Mart\'{\i}n for fruitful discussions.
E.M.-V.\ was financially supported by the Direcci\'on de Investigaciones of the Universidad Pedag\'ogica 
y Tecnol\'ogica de Colombia. A.I.F.-D. acknowledges funding from the Spanish Ministry of Economy and 
Competitiveness (MINECO, contract No.\ FIS2015-64951-R) and the EU Seventh Framework Programme under
Grant Agreement FP7-PEOPLE-2013-CIG-630996. J.C.C. thanks MINECO (contract No.\ FIS2017-84057-P) for 
financial support as well as the DFG and SFB 767 for sponsoring his stay at the University of
Konstanz as Mercator Fellow.

\appendix

\section{Additional results}

\begin{figure*}
\includegraphics[width=0.9\textwidth,clip]{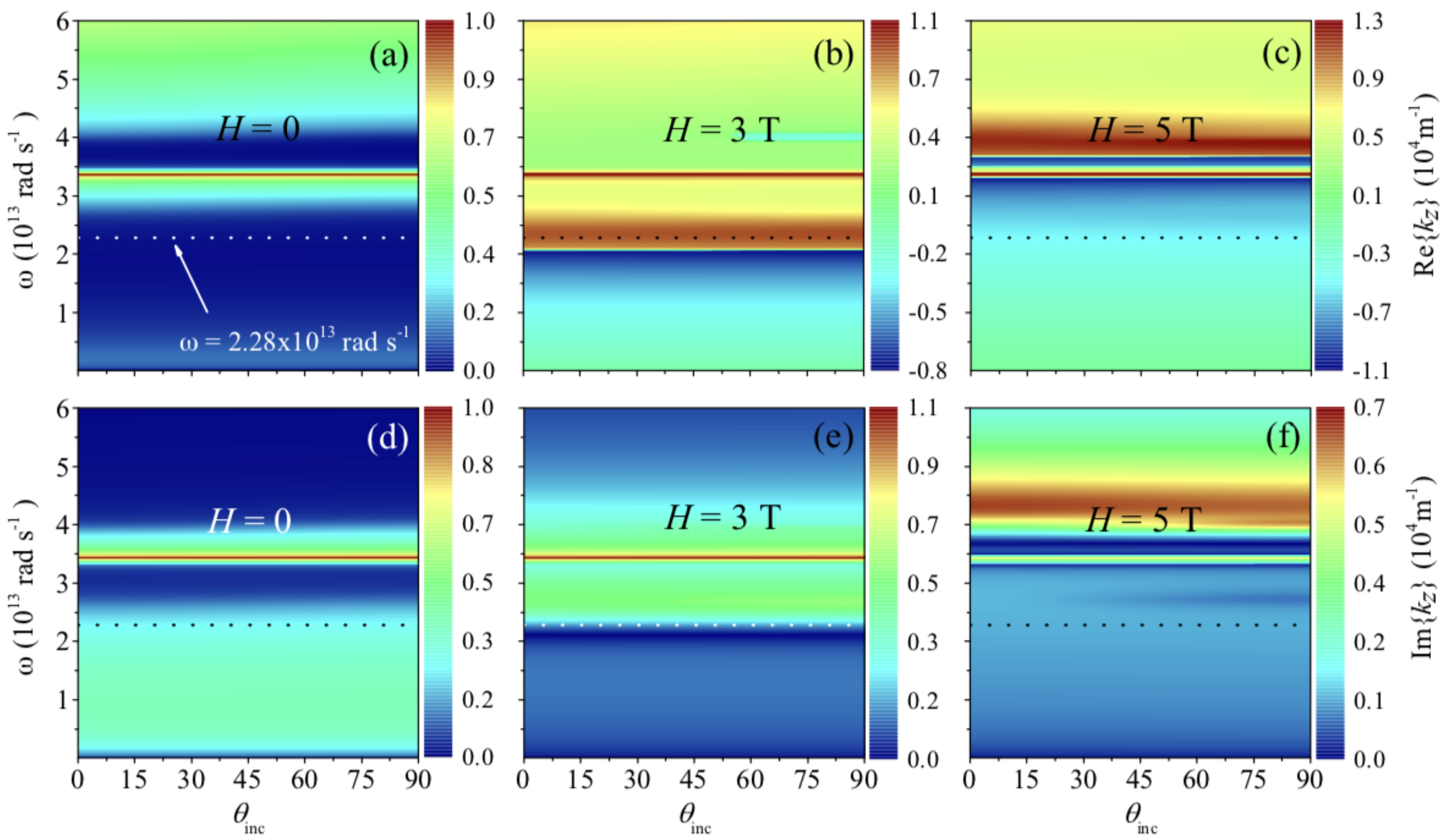}
\caption{(color online) Real (a)-(c) and imaginary (d)-(f) parts of $k_z$ for a TM 
polarized wave, see Eq.~\eqref{disptm}, as a function of the angular frequency and the angle of incidence 
for several values of the external magnetic field. The horizontal dotted lines indicate the same value of 
the frequency as in Figs.~\ref{fig_trans_H}(c,f).}
\label{fig-dispersion-ex}
\end{figure*}
\begin{figure*}
\includegraphics[width=0.9\textwidth,clip]{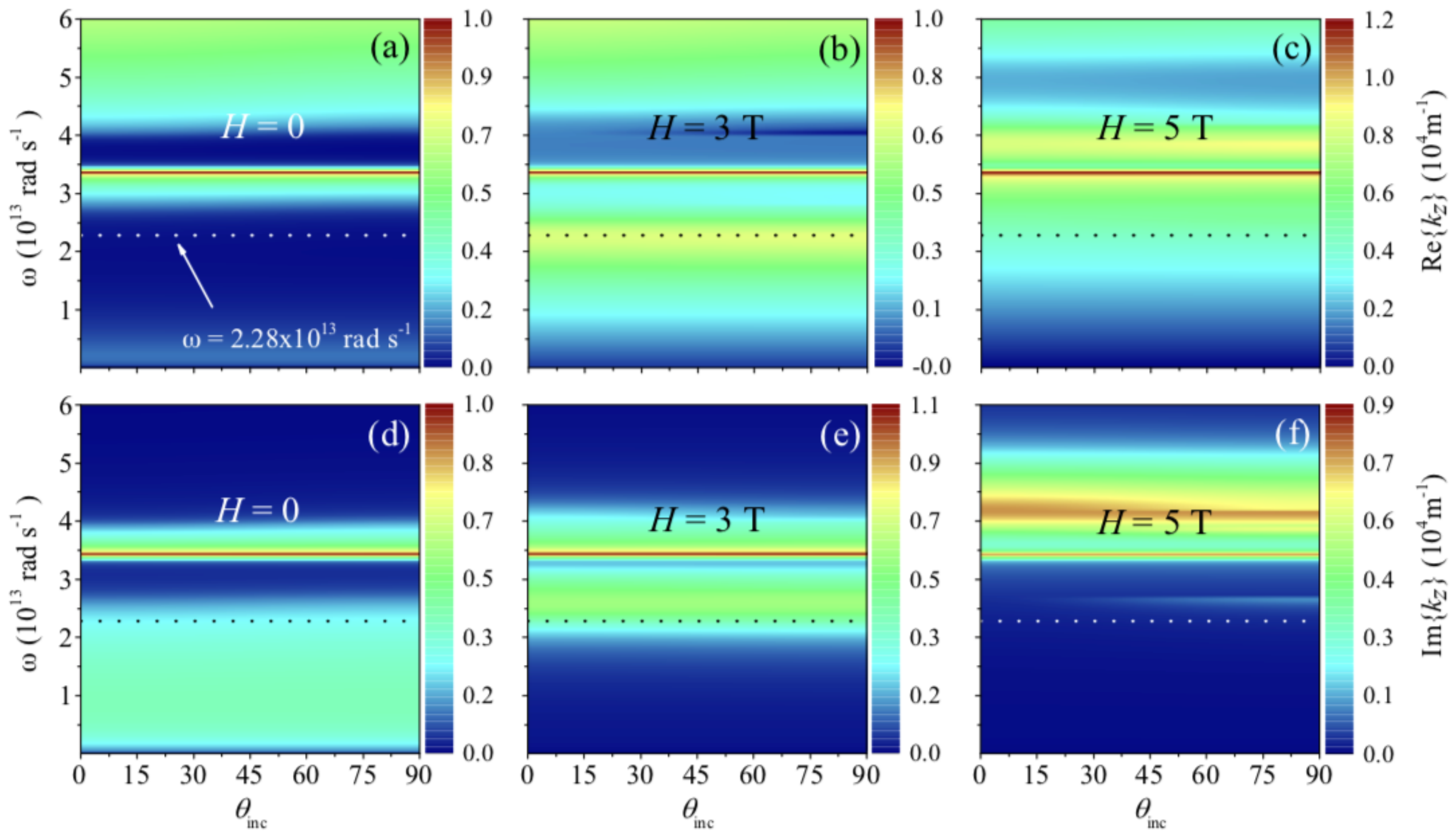}
\caption{(color online) The same as in Fig.~\ref{fig-dispersion-ex}, but with $\epsilon_{xy}=0$ 
(uniaxial approximation).}
\label{fig-dispersion-approx}
\end{figure*}

We present in this appendix additional results to clarify and back up some of the statements made in
the main body of the paper. First of all, and for the sake of completeness, we present here the 
exact dispersion relation of the propagating modes in the InSb slab under a magnetic field. Taking into
account the full structure of the permittivity tensor in Eq.~(\ref{perm-tensor-xy}), we obtain the
following dispersion relation for TE waves (propagating in the $xz$ plane)
\begin{widetext}
\begin{equation}
\label{dispte}
k_z^2 = \epsilon_{xx}\frac{\omega^2}{c^2}-\frac{\epsilon_{xx}+\epsilon_{zz}}
{2\epsilon_{zz}}k_x^2 - \sqrt{\left(\frac{\epsilon_{zz}-\epsilon_{xx}}{2\epsilon_{zz}}k_x^2\right)^2  
+ \frac{(\epsilon_{xy}k_x\omega)^2}{\epsilon_{zz} c^2} -  \epsilon_{xy}^2\frac{\omega^4}{c^4} },
\end{equation}
while for the TM waves this relation adopts the form
\begin{equation}
\label{disptm}
k_z^2= 	\epsilon_{xx}\frac{\omega^2}{c^2} - \frac{\epsilon_{xx}+\epsilon_{zz}}
{2\epsilon_{zz}}k_x^2+\sqrt{ \left(\frac{\epsilon_{zz}-\epsilon_{xx}}{2\epsilon_{zz}}k_x^2\right)^2  
+ \frac{(\epsilon_{xy}k_x\omega)^2}{\epsilon_{zz} c^2} -  \epsilon_{xy}^2 \frac{\omega^4}{c^4} } ,
\end{equation}
\end{widetext}
where $k_x = (\omega/c) \sin \theta_{\rm inc}$. These relations are rather involved. For this reason,
and in order to evaluate the impact of the off-diagonal elements of the permittivity tensor, we have 
plotted these full relations in Fig.~\ref{fig-dispersion-ex} for different values of the magnetic
field. In particular, we show both the real and the imaginary part of $k_z$ as a function 
of the frequency and the angle of incidence. To compare with the uniaxial approximation discussed in
section \ref{anom_refr}, we show in Fig.~\ref{fig-dispersion-approx} the corresponding results for the 
dispersion relation assuming that the off-diagonal elements vanish. As one can see, there are no 
dramatic differences, which explains in particular why the uniaxial approximation is able to describe 
the conditions at which the anomalous refraction occurs.

\begin{figure}[t]
\includegraphics[width=\columnwidth,clip]{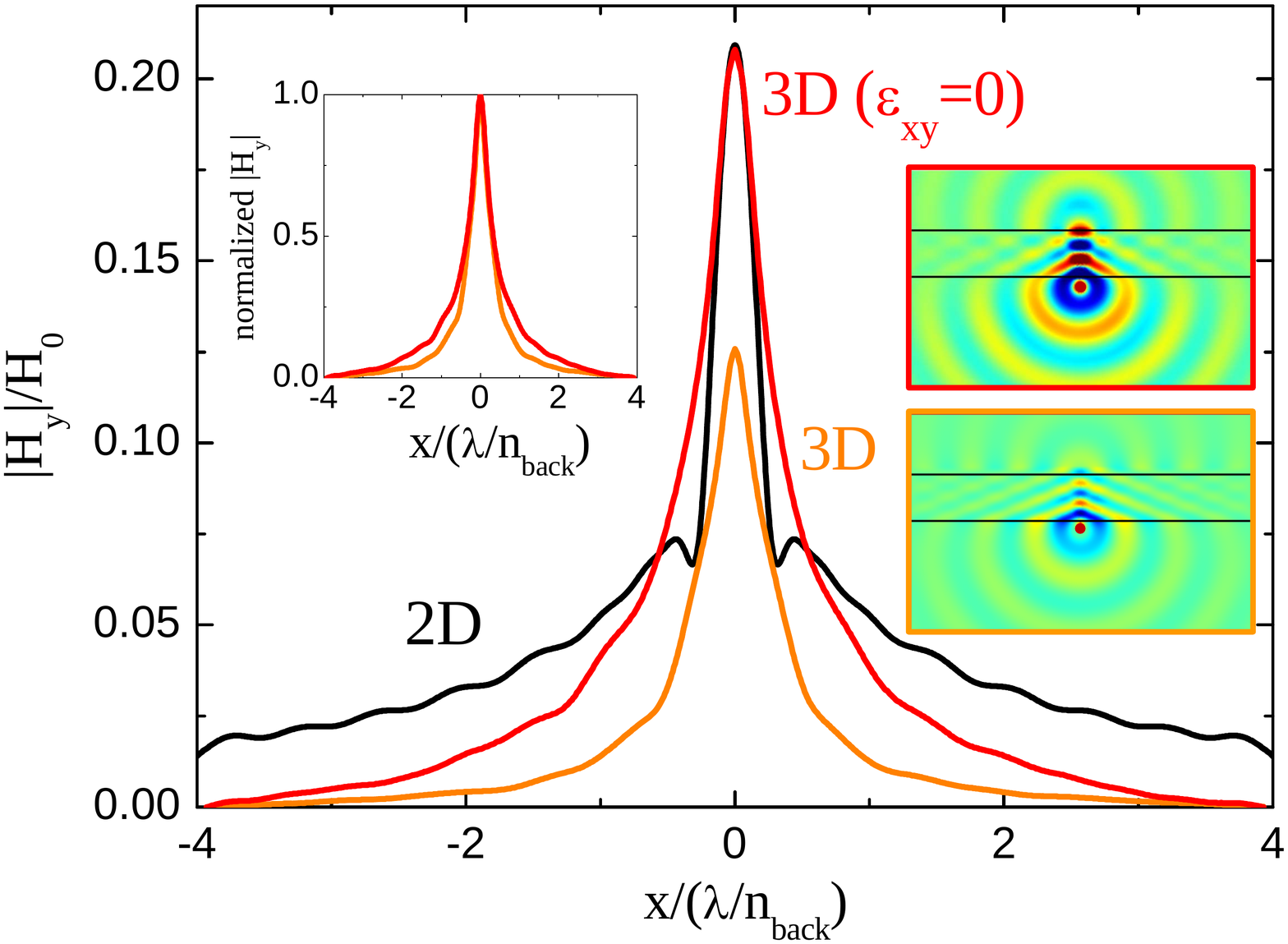}
\caption{(color online) The main panel shows the amplitude of the transmitted magnetic field normalized 
to $H_0$ (source field) evaluated at the surface of the slab in a simulation of the focusing when the 
external magnetic field is 5 T. The different parameters are identical to those of Fig.~\ref{fig_focusing}. 
The orange line corresponds to the full 3D simulation taking into account the off-diagonal elements
of the permittivity tensor, the red line is the 3D results without off-diagonal elements, and the 
black line corresponds to the 2D simulation. The left inset shows the two 3D results of the main panel,
but this time normalized to their maximum values. The right insets show the corresponding spatial distributions
of the $y$-component of the magnetic field in the 3D simulations with (orange frame) and without (red
frame) the off-diagonal components.} 
\label{fig_focusing_3D}
\end{figure}

Finally, in Fig.~\ref{fig_focusing_3D} we present the results corresponding to Fig.~\ref{fig_focusing}
for the focusing with an external field of 5 T, but this time computed with a full 3D simulation that 
takes into account the off-diagonal elements of the permittivity tensor. For comparison, we also include 
in this figure the results obtained with a 3D simulation without magneto-optics ($\epsilon_{xy} = 0$) and 
those of the 2D simulation of Fig.~\ref{fig_focusing}. Mimicking the 2D configuration, the source in the 3D
calculations is a line dipole along the $y$-direction. The convergence of results against the truncation of 
the simulation volume along this direction has been checked. As one can see, a significant focusing
survives in the full 3D simulations and the main characteristics of the field distribution are indeed
very similar to those reported in Fig.~\ref{fig_focusing} with the 2D approximation.


\end{document}